\documentclass[twocolumn,secnumarabic,amssymb, nobibnotes, aps, prx, superscriptaddress,longbibliography]{revtex4-2}

\usepackage{graphicx}% Include figure files
\usepackage{dcolumn}% Align table columns on decimal point
\usepackage{bm}% bold math
\usepackage{lipsum}
\usepackage[dvipsnames]{xcolor}
\usepackage[utf8]{inputenc}
\usepackage[T1]{fontenc}
\usepackage{mathptmx}
\usepackage{etoolbox}
\usepackage{tabularx}
\usepackage{xcolor}
\usepackage{chemformula}
\usepackage{siunitx}
\usepackage{booktabs}

\begin{document}	
	\title{Competing Ordinary and Hanle Magnetoresistance in Pt and Ti Thin Films}
	% Force line breaks with \\
	\author{Sebastian Sailler} \email{sebastian.sailler@uni-konstanz.de}
	\affiliation{Department of Physics, University of Konstanz, 78457 Konstanz, Germany}
	
	\author{Giacomo Sala}
	\affiliation{Department of Materials, ETH Zurich, H\"onggerbergring 64, 8093 Zurich, Switzerland}
	
	\author{Denise Reustlen}
	\affiliation{Department of Physics, University of Konstanz, 78457 Konstanz, Germany}
	
	\author{Richard Schlitz}
	\affiliation{Department of Physics, University of Konstanz, 78457 Konstanz, Germany}
	
	\author{Min-Gu Kang}
	\affiliation{Department of Materials, ETH Zurich, H\"onggerbergring 64, 8093 Zurich, Switzerland}
	
	\author{Pietro Gambardella}
	\affiliation{Department of Materials, ETH Zurich, H\"onggerbergring 64, 8093 Zurich, Switzerland}
	
	\author{Sebastian T. B. Goennenwein}
	\affiliation{Department of Physics, University of Konstanz, 78457 Konstanz, Germany}
	
	\author{Michaela Lammel} \email{michaela.lammel@uni-konstanz.de}
	\affiliation{Department of Physics, University of Konstanz, 78457 Konstanz, Germany}
	
	\date{September 20, 2024}% It is always \today, today,
	% but any date may be explicitly specified
	
	\begin{abstract}
		One of the key elements in spintronics research is the spin Hall effect, allowing to generate spin currents from charge currents. A large spin Hall effect is observed in materials with strong spin orbit coupling, e.g., Pt. Recent research suggests the existence of an orbital Hall effect, the orbital analogue to the spin Hall effect, which also arises in weakly spin orbit coupled materials like Ti, Mn or Cr. In Pt both effects are predicted to coexist. In any of these materials, a magnetic field perpendicular to the spin or orbital accumulation leads to additional Hanle dephasing and thereby the Hanle magnetoresistance (MR). To reveal the MR behavior of a material with both spin and orbital Hall effect, we thus study the MR of Pt thin films over a wide range of thicknesses. Careful evaluation shows that the MR of our textured samples is dominated by the ordinary MR rather than by the Hanle effect. We analyze the intrinsic properties of Pt films deposited by different groups and show that next to the resistivity also the structural properties of the film influence which MR dominates. We further show that this correlation can be found in both spin Hall active materials like Pt and orbital Hall active materials, like Ti. For both materials, the crystalline samples shows a MR attributed to the ordinary MR, whereas we find a large Hanle MR for the samples without apparent structural order. We then provide a set of rules to distinguish between the ordinary and the Hanle MR. We conclude that in all materials with a spin or orbital Hall effect the Hanle MR and the ordinary MR coexist and the purity and crystallinity of the thin film determine the dominating effect.
	\end{abstract}
	
	\maketitle
	
	\section{Introduction}
	
	Spintronics focuses on controlling and manipulating the spin degree of freedom, allowing to build promising devices for future applications, such as spin orbit torque magnetic random access memories or spin Hall nano-oscillators \cite{miron_perpendicular_2011, liu_spectral_2013, zhao_beyond_2015, chen_spin-torque_2016, han_spin-orbit_2021}. Here, spin orbit coupling (SOC) is an essential ingredient, as it connects the spin and charge degree of freedom. The spin Hall effect (SHE) leads to a spin current transversal to an applied charge current and vice versa a spin current generates a charge current via the inverse spin Hall effect (ISHE) \cite{wolf_spintronics_2001, kato_observation_2004, hoffmann_spin_2013, dyakonov_1971_spin, Hirsch_SHE, valenzula_nature_SHE, sinova_spin_hall_effects_2015}. For efficient devices, a large spin to charge conversion is desirable \cite{manchon_current_2019, krizakova_spin_orbit_2022}, making Pt the most commonly used material because of its large SOC and correspondingly large intrinsic SHE \cite{guo_intrinsic_2008}.
	
	% the spin Hall magnetoresistance (SMR)
	% The SMR describes the interplay of \textbf{s} with the magnetization \textbf{M} of an underlying ferromagnetic layer and is most prominently studied in nonmagnetic metal / ferrimagnetic insulator bilayers, such as yttrium iron garnet (YIG) / Pt \cite{nakayama_spin_2013, althammer_quantitative_2013, marmion_temperature_2014, wang_scaling_2014}. 
	
	Several magnetoresistances (MR), like the Hanle magnetoresistance (HMR) originate microscopically from an interplay of the SHE and ISHE, particularly from the resulting spin accumulation \textbf{s} \cite{dyakonov_magnetoresistance_2007, velez_hanle_2016}. The HMR arises from the interaction of the spin accumulation with the external magnetic field \textbf{B}. An increase in resistance is observed when the spin accumulation \textbf{s} includes a finite angle with \textbf{B}, which causes a torque on the spins, therefore a reduction of the spin accumulation and in turn the charge current converted by the ISHE \cite{dyakonov_magnetoresistance_2007,velez_hanle_2016}. %For both the SMR and the HMR, a large intrinsic spin Hall angle is crucial, as it directly influences the amplitude of both magnetoresistances \cite{chen_theory_2013, velez_hanle_2016}. 
	
	For the last decade most research focused on the spin degree of freedom (relying on materials with large SOC) \cite{wolf_spintronics_2001, zelse_spintronics_2004, sinova_spin_hall_effects_2015, chumak_magnon_2015}, while recently the orbital degree of freedom has gained significant interest, being reported in materials with weak SOC \cite{choi_observation_2023, lyalin_magneto_2023, sala_orbital_2023}. The microscopic description of the orbital Hall effect (OHE) is very similar to the one of the spin Hall effect, i.e. the OHE leads to an orbital accumulation \textbf{l}, which can then be observed or exploited \cite{bernevig_intrinsic_2008,go_intrinsic_2018}. The existence of the orbital Hall effect was experimentally confirmed in the light metals Ti and Cr \cite{choi_observation_2023, lyalin_magneto_2023} using optical methods as well as in Mn via a large orbital Hanle magnetoresistance \cite{sala_orbital_2023}.
	
	In some materials, e.g. Pt, the spin Hall as well as the orbital Hall effect occur simultaneously \cite{salemi_first_2022, sala_giant_2022}. However, no features in MR measurements that could be ascribed to the orbital Hall effect in Pt have been reported so far \cite{kumar_ultrafast_2023, moriya_observation_2024}. This might stem from the strong SOC in Pt, which causes one effective diffusion length to exist \cite{sala_giant_2022}. If Pt would show both effects separately, contributions stemming from the orbital Hall effect and from orbital to spin conversion might arise in addition to the known spin Hall contribution \cite{velez_hanle_2016}. MR experiments with different thicknesses could be a means to resolve this question.
	
	%If Pt would show both, a spin and an orbital Hall effect, two different diffusion lengths are expected, which should manifest as two maxima in the thickness dependence of the HMR.
	
	%This could be traced back to the orbital diffusion lengths as these are expected to be larger than the spin diffusion lengths. As the HMR shows a very distinct thickness dependence the influence of the orbital HMR could therefore be at higher thicknesses not investigated before. Additionally the orbital moment cannot interact with a substrates like YIG.
	%We expect two contributions for the HMR: one stemming from the spin and another one from the orbital degree of freedom. 
	%We show that next to the resistivity the respective diffusion coefficient plays a crucial role in determining the amplitude of the HMR.
	
	Here, we first investigate the MR of Pt thin films with various thicknesses on MgO substrates. We find a non-trivial and non-monotonous thickness dependence of our MR, which cannot be coherently explained within the framework of the HMR due to either the spin Hall or orbital Hall effect and which differs from other reports of the HMR in Pt thin films \cite{velez_hanle_2016, wu_hanle_2016, li_comprehensive_2022, sala_orbital_2023}. Upon careful analysis of our results, we find that the ordinary MR (OMR) \cite{pippard_magnetoresistance_1989} is the main effect contributing to the measured MR in our low resistivity samples. While the OMR can often be neglected in thin films, it was reported to occur simultaneously with spin Hall dependent effects \cite{isasa_spin_2016, li_comprehensive_2022}. We analyze how different material properties contribute to each MR and what causes the OMR or the HMR to be prevalent. We show that next to the resistivity the diffusion coefficient is a crucial parameter in determining the amplitude of the HMR, which in turn is heavily influenced by the crystallinity of the material. However, while the crystallinity influences the resistivity of the Pt, a scaling beyond the resistivity has to be taken into account to explain the differences in HMR amplitude from different groups. We then extend our study from Pt (as spin Hall active material \cite{velez_hanle_2016}) to Ti (as an orbital Hall active material \cite{choi_observation_2023}) and establish that the same mechanism and dependencies are valid for distinguishing the ordinary MR from the spin and/or orbital HMR. We show that also for Ti, a sizable HMR - which is most likely of orbital origin - can be found in samples with lower crystalline order.
			
	\section{Results and Discussion}

	\begin{figure}[t]
		\begin{center}
			\includegraphics[width=\linewidth]{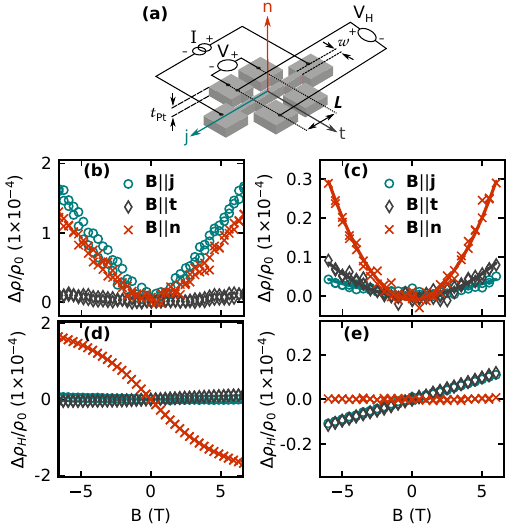}
			\caption{(a) Experimental setup and coordinate system utilized in this paper. After deposition of a Pt film with thickness $t_{\mathrm{Pt}}$, a Hallbar with a width $w$ and length between contacts $L$ is defined. A constant current $I$ is applied and the voltage drop $V$ is measured to determine the longitudinal magnetoresistance. Simultaneously, the transversal voltage V$_\mathrm{H}$ is detected. Along each of the coordinate axes in (a), the longitudinal resistivity $\rho (\textbf{B})$ and transversal resistivity $\rho_\mathrm{H} (\textbf{B})$ is calculated by taking the geometry of the Hall bar into consideration. (b) Change in longitudinal resistivity $\Delta \rho = \rho (\textbf{B}) - \rho_0(B\,=\,$\SI{0}{T}) normalized by $\rho_0$ of the exemplary MR for \SI{5}{nm} Pt as reported in Sala et al. \cite{sala_orbital_2023}, depicting a MR compatible with HMR theory, i.e., a finite MR for \textbf{B}$\perp$\textbf{t} and no MR for \textbf{B}||\textbf{t}. (c) Direction dependent MR for our \SI{8}{nm} thick Pt film as a function of the external field strength $B$, which cannot be explained by HMR theory. (d),(e) Transversal resistivity for the same samples as in (b) and (c). While a transversal HMR effect can be observed in the sample reported by Sala et al. \cite{sala_orbital_2023} (d), no extra signal can be found in our Pt (e). After subtraction of the linear ordinary Hall contribution along \textbf{B}||\textbf{n} (see Appendix~C), no change in transversal resistivity is found within the resolution limit.}
			\label{Fig_1_Setup}
		\end{center}
	\end{figure}
	
	To electrically characterize our Pt, we measure the MR in Hall bars with thicknesses ranging from \SI{2}{nm} to \SI{120}{nm}. For details regarding the fabrication please refer to Appendix~A. Fig.~\ref{Fig_1_Setup}(a) shows the experimental setup within the chosen coordinate system, where \textbf{j} is the axis along the current direction, \textbf{t} the axis transversal to it and \textbf{n} the axis parallel to the surface normal direction. Along these directions the resistivity $\rho$ is determined as a function of the external magnetic field. 
	
	Figure~\ref{Fig_1_Setup}(b) and (c) show the longitudinal MR $\frac{\Delta \rho}{\rho_0}$ versus \textbf{B}, where $\Delta\rho = \rho (\textbf{B}) - \rho_0$ is the change in longitudinal resistivity, $\rho_0$ the resistivity of Pt at \SI{0}{T}, for two different samples: a \SI{5}{nm} Pt thin film on \ch{SiO2} reported in Sala et al. \cite{sala_orbital_2023} [Fig.~\ref{Fig_1_Setup}(b)] and our \SI{8}{nm} Pt fabricated on MgO [Fig.~\ref{Fig_1_Setup}(c)]. As detailed in \cite{sala_orbital_2023}, the \SI{5}{nm} Pt shows a behavior consistently explained by HMR theory. Here, the relevant parameter is the spin accumulation $\textbf{s}$ (in our geometry \textbf{s}||\textbf{t}) and its direction with respect to the external magnetic field. If the magnetic field is perpendicular to \textbf{s}, Hanle dephasing causes the so called Hanle MR \cite{dyakonov_magnetoresistance_2007 ,velez_hanle_2016}, which can be seen for \textbf{B}||\textbf{j} and \textbf{n} [Fig.~\ref{Fig_1_Setup}(b)]. Along the \textbf{t} direction, no HMR is expected as \textbf{B}||\textbf{s}. For \textbf{B}||\textbf{j} and \textbf{n}, the HMR can be described via Eq.~\eqref{Eq_HMR}.
	
	\begin{equation}
		\frac{\Delta\rho}{\rho_0} = 2 \theta_\mathrm{s}^2 \left[\frac{\lambda_{\mathrm{s}}}{t_{\mathrm{Pt}}} 	\mathrm{tanh}\left(\frac{t_{\mathrm{Pt}}}{2\lambda_{\mathrm{s}}}\right) - \Re \left[\frac{\Lambda}{t_{\mathrm{Pt}}} \mathrm{tanh}\left(\frac{t_{\mathrm{Pt}}}{2\Lambda}\right)\right]\right]
		\label{Eq_HMR}
	\end{equation}
		
	Here, $t_\mathrm{Pt}$ is the film thickness, $\theta_\mathrm{s}$ the spin Hall angle, $\lambda_\mathrm{s}$ the spin diffusion length and $\Lambda^{-1}$ = $\sqrt{1/ \lambda^2 + i/\lambda^2_{\mathrm{m}}}$ with $\lambda_{\mathrm{m}}$ = $\sqrt{D \hbar / g \mu_B B}$ ($D$ = diffusion coefficient, $\hbar$ = reduced Planck constant, $g$ = gyromagnetic ratio, $\mu_B$ = Bohr magnetron) \cite{velez_hanle_2016}.
	
	In contrast to this expectation, our Pt shows a quadratic increase of the resistivity along all three directions [Fig.~\ref{Fig_1_Setup}(c)] and the MR amplitude increases from \textbf{j} to \textbf{t} to \textbf{n}. Furthermore, the MR is almost an order of magnitude smaller than the HMR in Sala et al. \cite{sala_orbital_2023} at the same magnetic field strengths.
	
	Additionally, the HMR also occurs in transversal geometry for \textbf{B}||\textbf{n} next to the ordinary Hall effect \cite{velez_hanle_2016, li_comprehensive_2022, sala_orbital_2023}, while along \textbf{j} and \textbf{t} neither the HMR nor the ordinary Hall effect are expected to occur. A transversal resistivity along \textbf{j} and \textbf{t} is therefore attributed to a misalignment with respect to \textbf{B}. To access the transversal HMR from the data for \textbf{B}||\textbf{n}, the linear contribution of the ordinary Hall effect of Pt is subtracted as detailed in Appendix~C. An emerging contribution from the HMR is then described by Eq.~\eqref{Eq_transverseHMR}:
	
	\begin{equation}
		\frac{\Delta \rho_2}{\rho_0} = 2 \theta^2_s \Im \left[\frac{\Lambda}{t_{\mathrm{Pt}}} \mathrm{tanh}\left(\frac{t_{\mathrm{Pt}}}{2\Lambda}\right)\right]
		\label{Eq_transverseHMR}
	\end{equation}
	
	Here, $\Delta \rho_2$ is the transversal resistivity stemming from the HMR while the other parameters are the same as in Eq.~\eqref{Eq_HMR} \cite{velez_hanle_2016}.
	
	While an HMR contribution to the transversal resistivity was found in the sample from Sala et al. \cite{sala_orbital_2023} [Fig.~\ref{Fig_1_Setup}(d)], no non-linear contribution can be seen in our Pt within the detection limit [Fig.~\ref{Fig_1_Setup}(e)]. While the transversal tanh($B$) like signal in Fig.~\ref{Fig_1_Setup}(d) is a strong corroboration for the longitudinal HMR of Sala et al. \cite{sala_orbital_2023}, the absence of it in our Pt [Fig.~\ref{Fig_1_Setup}(e)] does not allow for any conclusions, as the shape of the HMR contribution sensitively depends on $\lambda$ and $D$ and can take any form from linear to the tanh($B$) shape in Fig.~\ref{Fig_1_Setup}(d).

\begin{figure}[t]
		\begin{center}
			\includegraphics[width=\linewidth]{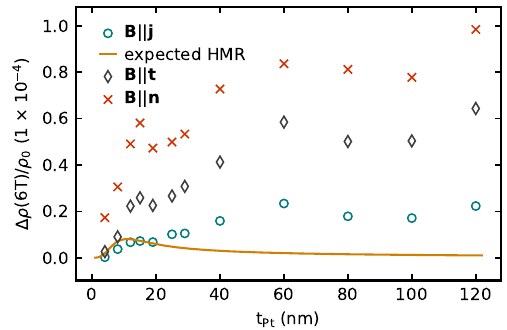}
			\caption{Normalized MR for the field sweeps along \textbf{j}, \textbf{t}, \textbf{n} as a function of Pt thickness $t_{\mathrm{Pt}}$. The amplitude $\Delta \rho$ at \SI{6}{T} is obtained from a quadratic fit to the data. In all cases, a non-trivial dependency on the thickness can be observed. The thickness dependence expected from the HMR theory with $\theta_\mathrm{s}$ = \SI{4}{\percent}, $\lambda_{\mathrm{s}}$ = \SI{2.4}{nm} (as calculated from Sagasta et al. \cite{sagasta_SHE}) and $D =$ \SI{3.1e-5}{m^2/s} (as discussed later [Fig.~\ref{Fig_5_params}(b)]) is shown in amber color. For thin films below \SI{20}{nm}, a reasonable agreement with the data for \textbf{B}||\textbf{j} is visible. Above that thickness, however, the expected HMR and the data deviate substantially. Furthermore, the changes in the MR are visible along all directions. From HMR theory no MR is expected for \textbf{B}||\textbf{t}, while similar amplitudes are expected for \textbf{B}||\textbf{j} and \textbf{B}||\textbf{n}.}
			\label{Fig_2_HMR_all}
		\end{center}
	\end{figure}

	To understand our MR, we start with analyzing the longitudinal HMR contribution. To that end, we evaluate the dependence of the HMR amplitude $\Delta \rho/ \rho_0$ on the film thickness, as a maximum of the HMR is expected at 4.56$\lambda$ \cite{velez_hanle_2016, li_comprehensive_2022}. Therefore, multiple samples were prepared, patterned and measured analogously to the sample in Fig.~\ref{Fig_1_Setup}(b). The amplitude of the MR $\Delta \rho / \rho_0$ is obtained by a quadratic fit to the data to extract $\Delta \rho$ at \SI{6}{T} for all field sweep directions and the results are shown in Fig.~\ref{Fig_2_HMR_all}. Here, a non-trivial thickness dependence is found which shows the same behavior along all three directions.

\begin{figure}[t]
		\begin{center}
			\includegraphics[width=\linewidth]{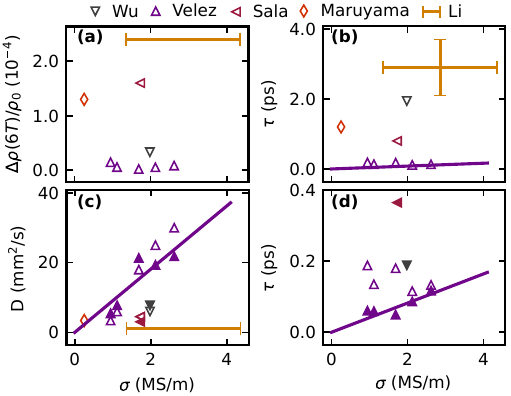}
			\caption{(a) HMR amplitude at \SI{6}{T}, (b) spin relaxation time $\tau$ for works from Tab.~\ref{Tab_MR}, (c) diffusion coefficient $D$ and (d) spin relaxation time $\tau$ for selected points on a smaller scale all versus the conductivity. It becomes apparent that for the HMR not only the resistivity but also the $D$ (or $\tau$) plays an important role. Since $\tau$ = $\lambda^2 / D$ and $\lambda \propto \sigma$ \cite{sagasta_SHE, liu_direct_2015}, a scaling with the conductivity is expected. From the Drude model we assume $\tau \propto \sigma$ and therefore also $D \propto \sigma$. The open symbols in (c) and (d) are the values as detailed in the respective work. While the linear fit is in reasonable agreement in (c), the data points in (d) deviate. For a better comparison, we recalculate all HMR parameters by taking the (intrinsic) scaling from Sagasta et al. \cite{sagasta_SHE} for $\sigma_\mathrm{s}$ and $\lambda_{\mathrm{s}}$ into account, which leaves only $D$ as the fit parameter. The so extracted $D$ is given by the full symbols. From this linear dependency, the diffusion coefficient used for HMR calculations in Fig.~\ref{Fig_2_HMR_all} and Fig.~\ref{Fig_A3_data_Hall} is calculated. The linear behavior shows, that the Pt from one group can consistently be described using the Drude assumption however, to explain the discrepancies between the different works, an additional contribution besides the conductivity (resistivity) is needed.}
			\label{Fig_5_params}
		\end{center}
	\end{figure}

	For \textbf{B}||\textbf{j}, typically only a contribution of the HMR as detailed above and no ordinary MR is expected. To approximate the value the HMR should exhibit in our Pt, we calculate the expected diffusion length $\lambda_\mathrm{s}$ and the expected (intrinsic) spin Hall angle $\theta_\mathrm{s}$ from the mean resistivity of all film thicknesses between \SI{4}{nm} and \SI{15}{nm} via $\lambda_\mathrm{s}\times \rho_0$ = \SI{0.61e-15}{\Omega m^2} and $\theta_\mathrm{s}/\rho_0$ = \SI{1.6e5}{\Omega^{-1}m^{-1}} as described by Sagasta et al. \cite{sagasta_SHE}. Using a mean resistivity of \SI{249}{n\Omega m} yields $\lambda_\mathrm{s}$ = \SI{2.4}{nm} and $\theta_\mathrm{s}$ = \SI{4}{\percent}. For the diffusion coefficient, we use $D =$ \SI{3.1e-5}{m^2/s}, extracted from a simple model as detailed further below [Fig.~\ref{Fig_5_params}]. The amber colored line in Fig.~\ref{Fig_2_HMR_all} shows the expected thickness dependence of the HMR for using above mentioned parameters. For films below \SI{20}{nm}, the calculation agrees reasonably with the MR along \textbf{j}, but deviates for thicker films, which suggests the existence of an additional contribution to the HMR next to the spin Hall physics. 
	 
	%Furthermore, the extracted parameters also explain the field dependency of the HMR shown in Fig.~\ref{Fig_1_Setup}(b), as for these parameters no saturation of the MR is expected for magnetic fields of up to \SI{6}{T}. The corresponding relaxation times $\tau$, extracted via $\tau$ = $\lambda^2/D$ are $\tau_\mathrm{s}$ = \SI{0.135}{ps} and $\tau_\mathrm{l}$ = \SI{0.141}{ps} for the spin and orbital parts, respectively. For $\tau$ values below \SI{1}{ps} a quadratic field dependence of the HMR is expected, which agrees with our experimental findings [see Fig.~\ref{Fig_1_Setup}(b)] \cite{velez_hanle_2016, li_comprehensive_2022}.
	%The differences between the two Pt films cannot be explained by the HMR alone, as no modulation is expected for applying the magnetic field along the \textbf{t} direction.
	
	Several implications arise from the assumption that the MR for \textbf{B}||\textbf{j} stems only from the HMR. First, one would expect the MR for \textbf{B}||\textbf{j} and \textbf{B}||\textbf{n} to have a similar amplitude. While the MR for \textbf{B}||\textbf{j} is mostly lower than the calculated value from the HMR theory, the MR along \textbf{n}, however, is much greater.
		
	Additionally, when analyzing the transversal component, no further contribution next to the ordinary Hall effect can be found, as detailed in Appendix~C. If the increase of the MR along \textbf{j} in the longitudinal data [Fig.~\ref{Fig_2_HMR_all}] were due to the Hanle effect, the absolute value of the transversal resistivity should deviate from the bulk ordinary Hall value. Since this is not the case we rule out a contribution of the Hanle MR for thicker films.
	
	Furthermore, the thickness dependencies for all three field directions [Fig.~\ref{Fig_2_HMR_all}] follow the same trend, which contradicts the expectation of the HMR theory where no MR is expected for \textbf{B}||\textbf{t}. The MR for \textbf{B}||\textbf{t} is usually attributed to the ordinary MR (OMR). The OMR arises when electrons are influenced by an external magnetic field $B$ in their respective cyclotron orbit. Depending on the orbit, the MR can saturate with increasing $B$ or exhibit a $B^i$ field dependence where $i \in [1,2]$ \cite{fawcett_high-field_1964, vasek_longitudinal_1975, pippard_magnetoresistance_1989}.
	
\begin{figure}[t]
		\begin{center}
			\includegraphics[width=\linewidth]{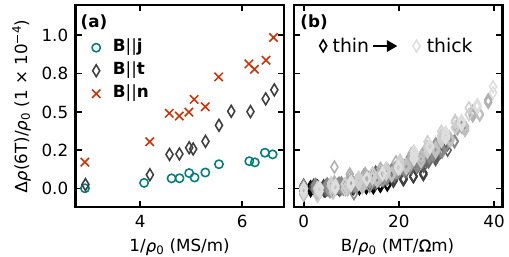}
			\caption{(a) Amplitude of the MR at \SI{6}{T} ($ \Delta \rho$) normalized by the resistivity at \SI{0}{T} ($\rho_0$) along all directions plotted versus $\rho^{-1}_0$ to highlight the scaling. (b) Kohler-plot of the MR for magnetic fields applied along \textbf{t}, where no HMR contribution is expected. The MR $ \Delta \rho / \rho_0$ is plotted over the external field divided by $\rho_0$. A color code from black (\SI{4}{nm}) to light gray (\SI{120}{nm}) is utilized to distinguish between the different thicknesses, which all show a similar behavior.}
			\label{Fig_3_Kohler}
		\end{center}
	\end{figure}

	To analyze the OMR contribution to our MR, we display the data in a so called Kohler-plot \cite{kohler_zur_1938, pippard_magnetoresistance_1989, isasa_spin_2016} [Fig.~\ref{Fig_3_Kohler}(b)]. There, the normalized amplitude of the magnetoresistance $\Delta\rho / \rho_0$ is plotted over $B$\,/\,$\rho_0$. This allows for a comparison of multiple samples regardless of their respective resistivities and at different temperatures \cite{kohler_zur_1938}. For the OMR, a scaling with $\rho^{-n}_0$ is expected, as described by Eq.~\eqref{Eq_Kohler}:
	
	\begin{equation}
		\frac{\Delta\rho}{\rho_0} = A_\mathrm{i} \left(\frac{B_\mathrm{i}}{\rho_0}\right)^n
		\label{Eq_Kohler}
	\end{equation}
	
	Here, $\frac{\Delta\rho}{\rho_0}$ is the OMR, $A_\mathrm{i}$ a material dependent factor with i = \textbf{j},\textbf{t},\textbf{n}, $B_\mathrm{i}$ the external magnetic field in the respective direction, and $\rho_0$ the resistivity at \SI{0}{T}. Experiments on various metals find exponents $n$ between 0 and 2 \cite{kohler_zur_1938, fawcett_high-field_1964, pippard_magnetoresistance_1989, isasa_spin_2016}. If the Kohler rule can be applied, it is a strong sign that the observed MR is the OMR.
	
	Fig.~\ref{Fig_3_Kohler}(a) depicts the scaling with $\rho^{-n}_0$ of the MR, whereas the Kohler plot of the MR data following Eq.~\eqref{Eq_Kohler} for the field sweeps along \textbf{t} is shown in Fig.~\ref{Fig_3_Kohler}(b). Here, no HMR is expected, as described above. A gray gradient is utilized to highlight the different samples, where black corresponds to the thinnest sample (\SI{4}{nm}) and the tone becomes increasingly brighter towards higher film thicknesses ($t_{\mathrm{max}}=$\SI{120}{nm}). Comparing the Kohler plots of all thicknesses shows that the Pt behaves similarly over all samples suggesting that the main MR contribution along \textbf{t} is the OMR. 

	 However, the MR along \textbf{n} in Fig.~\ref{Fig_2_HMR_all} still cannot be conclusively explained by taking the HMR extracted for \textbf{B}||\textbf{j} and the OMR extracted for \textbf{B}||\textbf{t} into account. There, the HMR and OMR contributions do not add up to yield the MR along \textbf{n}. 
	 
\begin{figure}[t]
		\begin{center}
			\includegraphics[width=\linewidth]{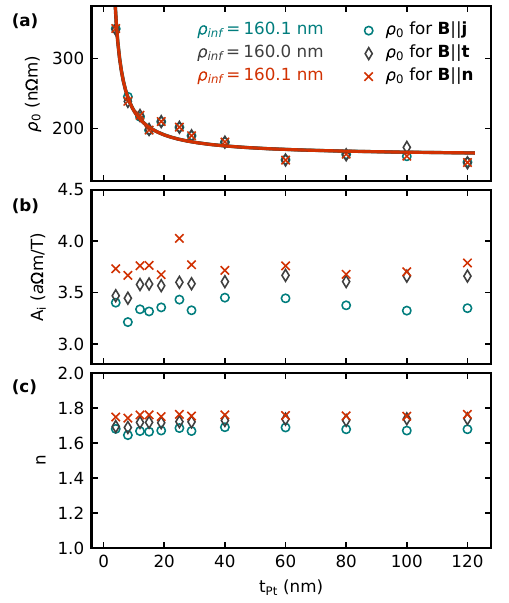}
			\caption{(a) Resistivity at \SI{0}{T} $\rho_0$, (b) amplitude $A_\mathrm{i}$ as well as (c) exponent $n$ of the respective Kohler fit versus the sample thickness. The resistivity for the bulk value is derived using an extended Fuchs-Sondheimer model \cite{fuchs_conductivity_1938, sondheimer_mean_1952, althammer_quantitative_2013}. The parameters obtained from fitting Eq.~\eqref{Eq_Kohler} to each individual sample along \textbf{j}, \textbf{t}, \textbf{n}, are almost identical for all samples and independent of the thickness. The mean values for $A_\mathrm{i}$ are 3.35, 3.58 and \SI{3.75}{a\Omega m \per T} and 1.67, 1.72 and 1.76 for $n$ along \textbf{j}, \textbf{t}, \textbf{n}, respectively.}
			\label{Fig_4_OMR}
		\end{center}
	\end{figure}

	 We therefore evaluate each individual field sweep along all directions utilizing Eq.~\eqref{Eq_Kohler} to investigate, if the OMR is the main mechanism in all directions. Fig.~\ref{Fig_4_OMR}(a) shows the resistivity, (b) the amplitude $A_\mathrm{i}$ and (c) the exponent $n$ of the Kohler fit to each individual sample over the thickness. The resistivity follows an extended Fuchs-Sondheimer Model \cite{fuchs_conductivity_1938, sondheimer_mean_1952, althammer_quantitative_2013} and the bulk resistivities $\rho_{\mathrm{inf}}$ of our Pt of \SI{160}{\nano\Omega\meter} are almost identical for each fit (see Appendix~B). The amplitudes $A_\mathrm{i}$ as well as the exponents $n$ are the lowest for field sweeps along \textbf{j}, increase to \textbf{t} and are the highest along \textbf{n}. The mean amplitudes are 3.35, 3.58 and \SI{3.75}{a\Omega m \per T} and the mean exponents are 1.67, 1.72 and 1.76, for \textbf{j}, \textbf{t}, \textbf{n}, respectively. Our extracted exponents are close to the $\mathrm{n} = 1.8$ reported before for Pt thin films \cite{isasa_spin_2016}. The values of $A_\mathrm{i}$ and $n$ for all sweep directions are basically independent of the thickness, which suggests that the MR along all three direction is the OMR.
	 
	The different amplitudes $A_i$ in Fig.~\ref{Fig_4_OMR}(b) for \textbf{B}||(\textbf{j}, \textbf{t}, \textbf{n}) demonstrate that the OMR can be anisotropic and most importantly also finite when the magnetic field is applied along the current, consistent with earlier studies \cite{vasek_longitudinal_1975, pippard_magnetoresistance_1989}. Consequently, a simple subtraction does not suffice when it occurs simultaneously with another MR. Furthermore, the crystallographic direction has a significant influence, as shown in crystal orientation dependent studies \cite{pippard_magnetoresistance_1989}. Only amorphous films or measurements along highly symmetric crystallographic directions (like the cubic $\langle$100$\rangle$ direction) are expected to show the same behavior. As our films grew with the $\langle$111$\rangle$ direction along the surface normal direction (see Appendix~A), the different orientations in the thin film plane are not equal, causing the amplitudes $A_\mathrm{i}$ and the exponents of the Kohler fit to differ [Fig.~\ref{Fig_4_OMR}(b), (c)].
	
	Figures~\ref{Fig_3_Kohler}(a) and \ref{Fig_4_OMR} further reveal that the observed MR $\Delta \rho / \rho_0$ in our Pt always scales with $\rho_0^{-n}$ and not the thickness as suggested by Fig.~\ref{Fig_2_HMR_all}. The resistivity $\rho_0$ is different for each sample and in turn depends on the thickness, explaining the dependence observed in Fig.~\ref{Fig_2_HMR_all}. The variations in the resistivity [Fig.~\ref{Fig_4_OMR}(a)] allow for a comprehensive explanation of the thickness dependence also for \textbf{B}||\textbf{j} [Fig.~\ref{Fig_2_HMR_all}]. In particular the low resistivity of the \SI{15}{nm} Pt film does correspond to a higher MR which leads to the apparent peak in the thickness dependence. Furthermore, the highest amplitudes of the MR in Fig.~\ref{Fig_2_HMR_all} are those of the samples with the lowest resistivity (i.e., the samples with $t_\mathrm{Pt} =$ \SI{60}{nm}, \SI{120}{nm}). This is consistent with the OMR theory, where fewer scattering processes lead to higher amplitudes, i.e., the purest samples show the highest OMR \cite{pippard_magnetoresistance_1989}. 
	
	For the HMR, the dependency with $\rho_0$ is more complicated \cite{velez_hanle_2016}. Taking the established (intrinsic) scaling of $\theta_{\mathrm{s}}$ $\propto$ $\rho_0$ and $\lambda_{\mathrm{s}} \propto 1/\rho_0$ \cite{sagasta_SHE} together with $D \propto 1/\rho_0$ [Fig.~\ref{Fig_5_params}(c)] leads to a minimal influence of the resistivity on the HMR amplitude. Mostly, it would only lead to a shift of the maximum HMR value towards higher (lower) film thicknesses for an decrease (increase) in resistivity, as the diffusion length would increase (decrease) according to $\lambda_\mathrm{s}\times \rho_0$ = \SI{0.61e-15}{\Omega m^2} \cite{sagasta_SHE}. However, to shift the maximum of the HMR towards film thicknesses of \SI{60}{nm}, where our maximum is located [Fig.~\ref{Fig_2_HMR_all}] a diffusion length of \SI{13.1}{nm} would be necessary. To obtain a $\lambda_\mathrm{s}$ of \SI{13.1}{nm} using the established $\lambda_\mathrm{s}\times \rho_0$ dependency, a resistivity below the bulk value of Pt would be required.
		
	Furthermore, as depicted by the solid line in Fig.~\ref{Fig_2_HMR_all}, the HMR amplitude is also expected to show a maximum and then decrease towards higher film thicknesses, which is not represented by the data. Together with the temperature dependence of the MR, exemplary shown for \textbf{B}||\textbf{n} in Appendix~D, this further corroborates the claim that no HMR is present in our Pt thin films.
	%As the prefactor in Eq.~\eqref{Eq_HMR} of $2 \theta^2$ scales with the resistivity, as $\theta \propto \rho_0$ \cite{wang_giant_2016, sagasta_SHE}, one would expect an increase in the MR with higher resistivity in the case of the HMR, opposed to the $\rho_0^{-n}$ scaling observed here [Fig.~\ref{Fig_3_Kohler}(a)]. 
	
	\begin{table}[b]
		\centering
		\newcolumntype{Y}{>{\centering\arraybackslash}X}
		\caption{Comparison of different studies on the HMR in Pt thin films. Depending on the resistivity of the Pt at zero field, either the OMR or the HMR dominate, with a range where both effects are observable.}
		\label{Tab_MR}
		\begin{tabularx}{\linewidth}{ Y Y Y}
			\hline \hline
			Ref & $\rho_0$ (n$\Omega$m) & MR \\
			\hline
			this work & 151-341 & OMR \\
			Li \cite{li_comprehensive_2022} & 200-500 & OMR + HMR \\
			Wu \cite{wu_hanle_2016} & 505 & HMR \\
			Sala \cite{sala_orbital_2023} & 580 & HMR \\
			Velez \cite{velez_hanle_2016} & 631-1059 & HMR \\ 
			Maruyama \cite{maruyama_modulation_2023} & 500 - 7000 & HMR \\
			\hline \hline
		\end{tabularx}
	\end{table}
	
	%However, when analyzing previously reported data it becomes evident that the resistivity is not the only parameter. To verify this claim, collected data of different groups is shown in Fig.~\ref{Fig_5_params} In the resistivity range multiple different HMR amp
	
	%One should note that the Kohler rule appears to be a good rule of thumb for differentiating between the MRs. When it is fulfilled, the Kohler scaling is a strong indication for an OMR. When it is not fulfilled, e.g. when one direction has no scaling with the external field, i.e. the exponent is not between 1 and 2, it is a strong indication of another MR. This is the case for the SMR, where the field dependence does not follow a simple linear or quadratic dependence. Furthermore, it also applies for the HMR (when the sample is amorphous), as the MR along \textbf{t} has no scaling with the field. (I think this is wrong, need to discuss next week). 
	%This train of thought can be extended. The Kohler rule is also not fulfilled for the anisotropic MR in ferrometals, where the MR is dependent on the direction of the 
	
	Despite the fact that Pt has a finite spin Hall effect, the MR of our Pt does not show any indications for an HMR caused by an orbital or spin accumulation. To reconcile this observation with the HMR picture, we use to two different material parameters: the resistivity $\rho_0$ and the diffusion coefficient $D$.
	
	First, our data suggest that the resistivity strongly influences whether the HMR or the OMR is dominant. So far, the HMR was exclusively observed for resistivity values larger than the ones reported here \cite{velez_hanle_2016, wu_hanle_2016, maruyama_modulation_2023, sala_orbital_2023}, which are summarized in Tab.~\ref{Tab_MR}. For the cleanest films (this study) no HMR is observed and for moderately dirty Pt ($\rho_0$ > \SI{500}{n \Omega m} \cite{velez_hanle_2016, wu_hanle_2016, maruyama_modulation_2023, sala_orbital_2023}) only the HMR is reported. For resistivities in the range between, both the HMR and the OMR are present \cite{li_comprehensive_2022}. We thus na\"ively expect that Pt thin films with a resistivity of approximately 5 times the bulk resistivity of $\rho_\mathrm{Pt}$~= \SI{105}{n\Omega m} \cite{rumble_CRC_2022}  exhibit a MR dominated by the HMR. 
	
	Although the resistivity of \SI{341}{n\Omega m} for our \SI{4}{nm} Pt film is in the range where both the HMR and the OMR have previously been observed \cite{li_comprehensive_2022} (see Tab.~\ref{Tab_MR}), we only find a response attributed to the OMR. This suggests that the resistivity cannot be the only factor determining the occurrence of the HMR \footnote{We can exclude that our Pt is not spin Hall active, as we observe a spin Hall magnetoresistance \cite{chen_theory_2013, nakayama_spin_2013} of $3.5\times10^{-4}$ in \SI{4}{nm} Pt (from the same sputtering process) on yttrium-iron-garnet (YIG). As both the spin Hall angle $\theta$ and the spin diffusion length $\lambda$ are part of HMR \cite{velez_hanle_2016} and SMR theory \cite{chen_theory_2013} and a function of the resistivity \cite{sagasta_SHE}, the diffusion coefficient $D$ must exert a significant influence on the HMR, as it is the only parameter different in the HMR as to the SMR}. Considering the dependencies for the spin Hall angle and spin diffusion length described by Sagasta et al. \cite{sagasta_SHE} ($\theta_\mathrm{s} \propto \rho_0$, $\lambda_\mathrm{s} \propto \rho_0^{-1}$) reveals that $D$ is the only free fit parameter in Eq.~\eqref{Eq_HMR} and further the only parameter where no direct scaling with the resistivity was reported. %Simply calculating the HMR at \SI{6}{T} for our \SI{4}{nm} thick sample with the diffusion coefficient from Velez et al. of $D =$ \SI{3.4e-6}{m^2/s} in \SI{3}{nm} Pt \cite{velez_hanle_2016} with $\theta_\mathrm{s} =$ \SI{5.5}{\percent} and $\lambda_{\mathrm{s}} =$ \SI{1.8}{nm} as derived from \cite{sagasta_SHE} would yield an HMR value of \SI{1.6e-4}{} opposed to the \SI{2.6e-6}{} measured here. This shows that $D$ hugely impacts the MR value and can differ even between Pt films.
	
	The diffusion coefficient directly influences the $\Re$ part of Eq.~\eqref{Eq_HMR} with an effect opposite to $B$, as  $\lambda_{\mathrm{m}} =\sqrt{D \hbar / g \mu_B B}$. An increase in $D$ increases the $\Re$ part of Eq.~\eqref{Eq_HMR} and therefore decreases the amplitude of the HMR. For a fixed $\lambda$, a fast (slow) diffusion implies a small (large) spin relaxation time and thus a weak (strong) interaction with $B$. Systems with spin relaxation times above $\approx$ \SI{1}{ps} then typically show the characteristic saturation towards high $B$ values in the longitudinal HMR and the tanh($B$) shape in the transversal MR \cite{li_comprehensive_2022}.
	
	To investigate the influence of $\rho_0$ and $D$ we compare the HMR from different groups in thin films between 2-\SI{7}{nm} [Fig.\ref{Fig_5_params}]. As the spin diffusion length is connected to the resistivity via $\lambda \propto \rho_0^{-1} \propto \sigma $ \cite{sagasta_SHE, liu_direct_2015} and connects $D$ and $\tau$ via $\tau$~= $\lambda^2 / D$, we expect $D$ and $\tau$ also to be a function of the resistivity. Fig.~\ref{Fig_5_params}(a) shows the MR amplitude at \SI{6}{T} for the samples given in Tab.~\ref{Tab_MR}, (c) the diffusion coefficient $D$ and (b),(d) the spin relaxation time $\tau$ versus the conductivity $\sigma = 1/\rho_0$. The open symbols in (c) and (d) are the values as taken from the papers from Tab.~\ref{Tab_MR}. As different assumptions were made depending on the work, we recalculated the parameters $\theta$ and $\lambda$ using the scaling with $\rho_0$ detailed in \cite{sagasta_SHE} and fit the HMR with only $D$ as a fitting parameters. The so extracted $D$ ($\tau$) is then marked by a full symbols in Fig.~\ref{Fig_5_params}(c),(d). Note that the data from Velez et al. \cite{velez_hanle_2016} are taken at \SI{100}{K} from Pt on \ch{SiO2} or Pyrex. 
	
	%We also include the values extracted from the spin HMR picture in Fig.~\ref{Fig_2_HMR_all} to indicate where our values would fall into the plot if they would be due to the HMR. The observed MR $\Delta \rho/\rho$ in our Pt is in good agreement with the values from Velez et al. \cite{velez_hanle_2016} and could therefore be falsely interpreted as an HMR. Since we concluded from the Kohler scaling that the HMR in our samples is much smaller than the OMR, the values reported here indicate the lower (upper) limit for $D$ ($\tau$) in our samples.
	 
	 %. These values ($\sigma$ = $1/\rho = 1/$\SI{249}{n\Omega m} = \SI{4}{MS/m}, $\lambda_\mathrm{s}$ = \SI{2.4}{nm}, $\theta_\mathrm{s}$ = \SI{4}{\percent} and $D_s$ = \SI{3.6e-5}{m^2/s}) are in good agreement with the values from Velez et al. \cite{velez_hanle_2016}. 
	%The parameters extracted here via HMR theory from Fig.~\ref{Fig_2_HMR_all} are interestingly in good agreement with Velez et al. \cite{velez_hanle_2016}. This could easily lead to a false interpretation as the MR amplitudes are in the same range. However, as we do not observe the HMR, the values for $D$ ($\tau$) only serve as lower (upper) limits. 
	
	From Fig.~\ref{Fig_5_params} it becomes evident that a simultaneous description for $D$ or $\tau$ is difficult and that a scaling beyond the resistivity has to be considered. For further evaluation we take the values extracted from Velez et al. \cite{velez_hanle_2016} as multiple samples were tested there and then use the Drude model as a starting point with $\sigma \propto \tau$. With the previously mentioned dependencies of $\tau$~=~$\lambda^2 / D$ and $\lambda \propto \sigma$ this leads to a linear scaling of both $D$ and $\tau$ with $\sigma$, which is consistent with the Einstein relation where $D \propto \tau$. A linear fit to $D$ (full symbols) versus $\sigma$ in Fig.~\ref{Fig_5_params}(c) yields a slope of \SI{9.06e-12}{\Omega m^3 s^{-1}}. Together with $\lambda / \sigma$ = \SI{0.61e-15}{\Omega m^2} \cite{sagasta_SHE} we obtain $\tau/\sigma$ = \SI{4.1e-20}{\Omega ms}, i.e., the linear slope shown in Fig.~\ref{Fig_5_params}(b) (and on a smaller scale for better readability in Fig.~\ref{Fig_5_params}(d)). This shows that all Pt thin films from Velez et al. \cite{velez_hanle_2016} can consistently be explained with a linear scaling of both $\tau$ and $D$ in $\sigma$.
	
	%Furthermore, the values extracted here for $D$ and $\tau$ are in good agreement with our Pt and would fit nicely into the data set of \cite{velez_hanle_2016}. This means, that generally our values would be consistent in an HMR picture, but the real value for $D$ is higher (or $\tau$ smaller) as we do not see the HMR. Another interpretation could be, that the HMR is part of the OMR and contributes to the MR||\textbf{j} and \textbf{n}.
	
	However, the values of $\tau$ ($D$) from other reports shown in Fig.~\ref{Fig_5_params} do not coincide with the $\sigma$ scaling extracted from the data by Velez et al. \cite{velez_hanle_2016}, differing by orders of magnitude. To exclude roughness or thickness variations, one would need to compare the $\rho_{\mathrm{inf}}$ value extracted from the extended Fuchs-Sondheimer model \cite{fuchs_conductivity_1938, sondheimer_mean_1952, althammer_quantitative_2013}, which we only have for our own films. Additionally, a thickness or roughness variation would again lead to a scaling in $\rho_0$ (or $\sigma$) and could neither explain the difference in $D$ (or $\tau$) nor the amplitude of the HMR. We thus conclude that there must be another contribution to the diffusion coefficient which is not directly included in the resistivity.
		
	We propose that this additional contribution to $D$ stems from the crystalline quality of the Pt thin film. Although the same deposition technique and the same substrates were used here and in \cite{velez_hanle_2016, wu_hanle_2016, sala_orbital_2023}, the MR behaves very different [Fig.~\ref{Fig_5_params}(a),(b),(c)], rendering a direct dependency from the substrate or the deposition technique unlikely. Although the deposition and the substrate might not directly influence the HMR, both can affect the crystal quality of the Pt. This in turn influences $D$, as both the diffusing species as well as the host material influences its value (at a given temperature) \cite{callister_materials_2007}. The larger HMR amplitude that is reported by some groups could also stem from an additional contribution from the orbital Hall effect. As spin and orbit are strongly coupled in Pt, only one effective $\lambda$ is expected, which could also not be differentiated in the HMR \cite{sala_giant_2022}. A further possibility is the existence of surface states with a low (high) D ($\tau$), effectively trapping the electrons \cite{rouzegar2023terahertz}.

	\begin{figure*}[t]
		\begin{center}
			\includegraphics[width=\linewidth]{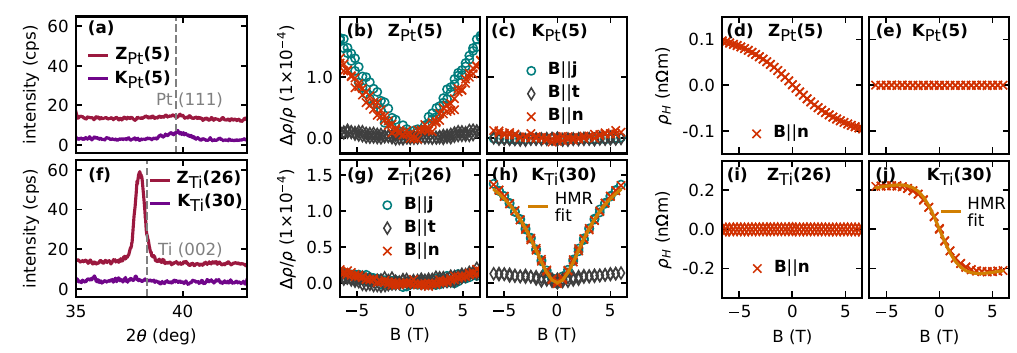}
			\caption{Top row: Pt characterization. (a) X-ray diffraction (XRD) data from a symmetrical $\theta-2\theta$ scan of \SI{5}{nm} thick Pt film on \ch{SiO2} from Konstanz K$_\mathrm{Pt}$(5) and Z\"urich Z$_\mathrm{Pt}$(5). For both samples a small increase in counts per second (cps) is visible at the expected $\langle$111$\rangle$ peak position of Pt, with the peak from K$_\mathrm{Pt}$(5) being more pronounced. (b), (c) Longitudinal MR data for \textbf{B}||\textbf{j},\textbf{t},\textbf{n} in \SI{5}{nm} Pt made in Z\"urich (Z$_\mathrm{Pt}$) and Konstanz (K$_\mathrm{Pt}$). (d), (e) Transversal resistivity after subtraction of a linear slope from the ordinary Hall effect (see Appendix~C). Only the data for \textbf{B}||\textbf{n} is shown, as Hall type effects are sensitive to this direction. The less crystalline sample Z$_\mathrm{Pt}$(5) shows a MR consistently explained in the HMR picture as described in \cite{sala_orbital_2023}, while the more crystalline K$_\mathrm{Pt}$(5) can be described in the OMR picture. This gets supported by the transversal resistivity, where no remaining signal is found for the OMR, while a non-linear behavior is seen for the HMR. \linebreak Bottom row: Ti characterization. (f) XRD data for Ti from Konstanz and Z\"urich with thicknesses of 26 and \SI{30}{nm}. Here, the difference in crystallinity becomes clearer as in Pt. (g) - (j) Same as in (b) - (e) just for Ti: The crystalline sample shows a small MR similar to the OMR picture in Pt while the amorphous sample K$_\mathrm{Ti}$(30) exhibits a saturation in the longitudinal MR as well as a tanh($B$) shape in the transversal resistivity, consistent with HMR theory.}
			\label{Fig_6_pt_ti}
		\end{center}
	\end{figure*}
	
	The crystal quality of the Pt thin films, as determined via X-ray diffraction (XRD), differs between the groups, depends on the substrates and will be discussed in the following. While Velez et al. report barely any intensity for their Pt peak, our Pt on MgO becomes highly crystalline, showing Laue osciallations on the $\langle$111$\rangle$ peak of Pt (see Appendix~A, Fig.~\ref{Fig_A1_XRD}). To further investigate how the crystallinity influences the MR and for better comparison with other studies we prepared two samples of \SI{5}{nm} Pt on a thermally oxidized Si/\ch{SiO2} wafer (K$_\mathrm{Pt}$(5)) and compare it with the Pt sample (Z$_\mathrm{Pt}$(5)) made at the ETH Z\"urich shown in Fig.~\ref{Fig_1_Setup}(b) and Fig.~\ref{Fig_5_params} from \cite{sala_orbital_2023}. The samples are labeled with their respective place of fabrication by sputtering (\textbf{K}onstanz vs \textbf{Z}\"urich) with the process details being described in Appendix~A. The magnetotransport responses of the two nominally identical Pt films are fundamentally different, as shown in Fig.~\ref{Fig_6_pt_ti}. 
	
	To access the crystallinity of the Pt layers, Fig.~\ref{Fig_6_pt_ti}(a) shows the intensity versus the $2\theta$ angle from a symmetrical $\theta - 2\theta$ XRD scan. For better readability the two XRD spectra are shifted with respect to each other. The intensity increase around the expected $\langle$111$\rangle$ Bragg peak of Pt \cite{narayan_formation_1994} is more pronounced for K$_\mathrm{Pt}$(5), signalizing a better crystalline quality of this Pt layer.
	
	The electrical properties of the two different \SI{5}{nm} Pt thin films on \ch{SiO2} (K$_\mathrm{Pt}$(5) and Z$_\mathrm{Pt}$(5)) are shown in Fig.~\ref{Fig_6_pt_ti}(b)-(e). K$_\mathrm{Pt}$(5) has a $\rho_0$ of \SI{421}{n\Omega m} and shows a longitudinal MR with an amplitude for \textbf{B}||\textbf{j}~< \SI{1e-6}{} at \SI{6}{T} with the MR for \textbf{B}||\textbf{n} greater than \textbf{B}||\textbf{j} and \textbf{B}||\textbf{t} [Fig.~\ref{Fig_6_pt_ti}(c)]. The behavior is comparable to the samples on MgO, which we ascribe to a dominating OMR. In contrast to that, the sample Z$_\mathrm{Pt}$(5) with $\rho_0$ of \SI{580}{n\Omega m} can be consistently explained by the HMR as described in \cite{sala_orbital_2023} [Fig.~\ref{Fig_6_pt_ti}(b)]. This is corroborated by the corresponding transversal signal. After subtraction of the linear ordinary Hall effect (see Appendix~C, Fig.~\ref{Fig_A2_transervsesetup}), the underlying signal is shown in Fig.~\ref{Fig_6_pt_ti}(d) and (e). For K$_\mathrm{Pt}$(5), no additional contribution within the resolution limit can be detected [Fig.~\ref{Fig_6_pt_ti}(e)]. In comparison, Z$_\mathrm{Pt}$(5) shows a non-linear residue [Fig.~\ref{Fig_6_pt_ti}(d)], compatible with the Hall type contribution of the HMR, as described by Eq.~\eqref{Eq_transverseHMR} (see Appendix~C).
	
	Despite the similar resistivities and thicknesses, as well as the same substrate and deposition technique, the prominent MR and the order of magnitude of the MR amplitude vary significantly as seen in Fig.~\ref{Fig_6_pt_ti}(b)-(e). Since our Pt thin films are more consistently explained within the OMR picture, we propose that a higher crystallinity and crystal quality of the samples leads to a higher diffusion coefficient. This dependency is corroborated by comparing the $D$ values reported in the literature for Pt of $1 \times 10^{-6}$ to \SI{6e-6}{m^2/s} \cite{velez_hanle_2016, li_comprehensive_2022, sala_orbital_2023} with the expected $D$ for our Pt of \SI{3.1e-5}{m^2/s}. Note that this is the value expected from the scaling depicted in Fig.~\ref{Fig_5_params}(c). The value for $D$ only serves as a lower limit, as we rule out a dominant HMR contribution. This observation is consistent with our findings and with HMR theory, which predicts that the HMR is less prevalent in samples with a higher $D$.
	
	Despite these apparent discrepancies, as a general rule of thumb a resistivity 5 times the bulk value of Pt should lead to a discernible HMR response. Note, however, that the resistivity alone is not sufficient to distinguish between the two MRs. The effect size of the HMR is largely influenced by the diffusion coefficient, meaning that very fast diffusion can lead to low HMR values even for high resistivity films.
			
	To further corroborate the importance of the crystallinity, we analyze Ti samples of similar thicknesses also from Konstanz and Z\"urich. In Ti the orbital Hall effect was first experimentally observed via optical measurements \cite{choi_observation_2023}. We compare samples with a Ti thickness of $t_\mathrm{Ti}~=$ \SI{26}{nm} from Z\"urich (Z$_\mathrm{Ti}$(26)) with a samples from Konstanz $t_\mathrm{Ti}~=$ \SI{30}{nm} (K$_\mathrm{Ti}$(30)), both deposited on Si/\ch{SiO2} wafers.
	
	The structural and electrical properties of the different Ti thin films are depicted in Fig.~\ref{Fig_6_pt_ti}(f)-(j). A clear difference in crystallinity between the two films becomes apparent from the XRD spectra [Fig.~\ref{Fig_6_pt_ti}(f)]. Although the samples are of similar thickness, a clear intensity increase near the nominal Ti $\langle$002$\rangle$ position can only be observed for Z$_\mathrm{Ti}$(26), signalizing a crystalline thin film, whereas K$_\mathrm{Ti}$(30) still appears to be amorphous. Correlating these results to the transport data shown in Fig.~\ref{Fig_6_pt_ti}(g)-(f), again corroborates that the crystallinity is a key factor influencing the dominating MR. 
	
	The longitudinal MR of the crystalline sample Z$_\mathrm{Ti}$(26) with $\rho_0$(Z$_\mathrm{Ti}$(26))~= \SI{1.31}{\mu\Omega m} is almost identical along all directions with an amplitude of $\Delta\rho/\rho_0 \approx$ \SI{15e-6}{} at \SI{6}{T}, consistent with the OMR picture. This is supported by the transversal data in in Fig.~\ref{Fig_6_pt_ti}(i), where no additional Hall type contribution is found after subtraction of the linear ordinary Hall effect (see Appendix~C).
	
	The longitudinal MR of the amorphous sample K$_\mathrm{Ti}$(30) ($\rho_0$(K$_\mathrm{Ti}$(30))~= \SI{2.49}{\mu\Omega m}) on the other hand suggests a saturation towards sufficiently high magnetic fields. For the Ti samples from Konstanz, the MR along \textbf{t} is small, with an amplitude in the same range as the MR in the crystalline sample Z$_\mathrm{Ti}$(26), which is consistent with the OMR there. Note that the amplitude of the MR at \SI{6}{T} is almost 10 times higher than the one of Z$_\mathrm{Ti}$(26) [Fig.~\ref{Fig_6_pt_ti}(h)]. Furthermore, a pronounced tanh($B$) shaped residue is found after subtraction of the ordinary Hall effect [Fig.~\ref{Fig_6_pt_ti}(j)], which suggests that the arising MR is dominated by the HMR. As in Ti spin contributions are predicted to be negligible, we attribute the effect to stem from an orbital contribution \cite{salemi_first_2022}.
	
	To analyze the orbital contribution, a fit to the MR data of K$_\mathrm{Ti}$(30) is conducted for the longitudinal MR according to Eq.~\eqref{Eq_HMR} and for the transversal component according to Eq.~\eqref{Eq_transverseHMR}. A simultaneous fit of both data sets yields $\theta_{\mathrm{l}} =$ \SI{2.3}{\percent}, $\lambda_{\mathrm{l}} =$ \SI{7.0}{nm} and $D =$ \SI{1.8e-5}{m^2/s}, with the fit being depicted in amber color in the respective Fig.~\ref{Fig_6_pt_ti}(h) and (j). The extracted diffusion length of \SI{7.0}{nm} lies well below the 50 to \SI{60}{nm} estimated in Choi et al. \cite{choi_observation_2023}, but well above typical diffusion lengths in Pt of 1-\SI{2}{nm} \cite{velez_hanle_2016, sala_orbital_2023}. Performing the calculations for the HMR with these diffusion lengths as fixed parameters allows to determine an upper (\SI{4.3}{\percent} for $\lambda_{\mathrm{l}} =$ \SI{2}{nm}) and lower (\SI{1.5}{\percent} for $\lambda_{\mathrm{l}} =$ \SI{60}{nm}) boundary for the (orbital) Hall angle in Ti. These Hall angles would correlate to an intrinsic orbital Hall conductivity $\sigma_{\mathrm{OH}} = (\hbar/e) \theta_{\mathrm{l}} / \rho_{\mathrm{Ti}} = $ \SI{17403}{(\textit{$\hbar /$e}) (\Omega m)^{-1}} for $\lambda_{\mathrm{l}} =$ \SI{2}{nm} ($\sigma_{\mathrm{OH}} =$ \SI{5957}{(\textit{$\hbar /$e}) (\Omega m)^{-1}} for $\lambda_{\mathrm{l}} =$ \SI{60}{nm}), in line with experimental results \cite{choi_observation_2023}, but well below theoretical values \cite{salemi_first_2022}.
	
	We therefore conclude that the MR in K$_\mathrm{Ti}$(30) is dominated by the Hanle effect, whereas the MR in Z$_\mathrm{Ti}$(26) is governed by the OMR. The resistivites of the individual samples of $\rho_0$(K$_\mathrm{Ti}$(30))~= \SI{2.49}{\mu\Omega m} and $\rho_0$(Z$_\mathrm{Ti}$(26))~= \SI{1.31}{\mu\Omega m} also fit into our proposed regimes with respect to the resistivities, i.e., K$_\mathrm{Ti}$(30)$ > 5\times\rho_{0,\mathrm{Ti}} >$~Z$_\mathrm{Ti}$(26), with $\rho_{0,\mathrm{Ti}} =$ \SI{450}{n\Omega m}. 
	
	Note that while the resistivity can be used as indicator for the expected regime, it cannot explain the quantitative differences in the MR amplitudes. Both data sets, Ti and Pt, suggest that the diffusion coefficient becomes larger with increasing crystallinity of the metal layer. However, the increasing $D$ is not captured by the Drude model as this only accounts for the "bulk" electronic properties. Especially for the theoretical description of the orbital contribution, crystalline order plays an important role \cite{go_intrinsic_2018, tang2024roledisorderintrinsicorbital}. Interestingly, sample K$_\mathrm{Ti}$(30) [Fig.~\ref{Fig_6_pt_ti}(h) and (j)] shows a sizable HMR effect despite the lack of crystalline order, which we tentatively ascribe to the orbital Hall effect in Ti. However, this would indicate that the orbital accumulation is conserved even without a defined crystal structure.
	
	Under the assumption that the orbital diffusion resembles the spin diffusion, we expect a similar scaling for $\theta$ and $\lambda$ in Ti as for Pt. An increase in $\lambda$ is accompanied with a decrease in $\theta$. This dependency together with a larger $D$ in the crystalline sample could explain the quantitative differences between the two samples. In the crystalline sample we expect robust orbital accumulation as predicted by theory and because of a fast $D$, little interaction with $B$. This explains, why the HMR contribution is small and below the measured MR in Fig.~\ref{Fig_6_pt_ti}(g),(i).
		
	For both materials, Pt and Ti, a large intrinsic spin/orbital Hall effect does not guarantee the occurrence of the HMR. Instead, it is important to also take the crystallinity of the film into account. This consequently means that the type of scattering within the metal is of importance and greatly influences the resulting properties, which reflects in recent theoretical discussion regarding the importance of defects in the lattice \cite{tang2024roledisorderintrinsicorbital, liu_dominance_2024}.
	
	%more crystalline means longer diffusion length and lower angle. For the similar resistivities in K$_\mathrm{Pt}$(5) and Z$_\mathrm{Pt}$(5) this means that even the type of scattering is important. An interesting observation is that for the theoretical description - especially for the orbital Hall effect \cite{go_intrinsic_2018} - a clean band structure and defined orbits are necessary, whereas experimentally, the opposite seems to be ideal to observe a sizable effect.}

	%Even though Pt has a large intrinsic spin Hall effect \cite{guo_intrinsic_2008} and both Ti and Pt have a sizable orbital Hall effect \cite{salemi_first_2022} the HMR cannot always be observed. This does not mean, that the HMR does not exist in the respective material, only that the amplitude becomes very small and at some point becomes smaller than the OMR. This happens when the material becomes to "clean", meaning few impurities and a low resistivity. Additionally, for the observation of the HMR a low diffusion coefficient is necessary, leading to an accumulation of spins or orbital moments on which the external field can act upon. For the same material we suggest a slower diffusion coefficient when the material is amorphous. 
	
	We propose that the HMR and OMR coexist in all metals with a spin (or orbital) Hall conductivity. Solely observing one does not mean that the other MR does not exist in the respective material. Which one of the two effects dominates the MR sensitively depends on the material parameters. While in clean, crystalline materials (few impurities, low resistivity, high diffusion coefficient) the OMR is more prevalent, the HMR is expected to govern the MR in samples which are amorphous and sufficiently dirty (more impurities, higher resistivity, low diffusion coefficient).
	
	Experimentally, the differentiation between the HMR and the OMR is not always straight forward. For the longitudinal data, similar MR amplitudes for \textbf{B}||\textbf{j} and \textbf{n} and a significantly smaller MR for \textbf{B}||\textbf{t} are strong indicators for the HMR. For the OMR, the MR for \textbf{B} along \textbf{j} is typically smaller than for \textbf{B} perpendicular to \textbf{j} \cite{pippard_magnetoresistance_1989, nickel_magnetoresistance_overview_1995}. In a Kohler plot, similar pre-factors $A_i$ and exponents $n$ are expected along all sweep directions. To rule out the influence of crystalline anisotropies on the OMR, the MR should either be investigated along highly symmetric directions or in amorphous materials. The OMR is then characterized by a scaling with $\rho^{-n}$, whereas the HMR exhibits a characteristic scaling with the sample thickness. Furthermore, observing a tanh($B$) shape in the transversal data in addition to the ordinary Hall effect [Fig.~\ref{Fig_6_pt_ti}(d) and (j)] can be a strong sign of the HMR. Yet, it is not sufficient alone, as for samples with a high $D$, the curve shape can be linear and therefore hard to distinguish from the ordinary Hall effect. Moreover, the non-linear shape can also be caused by multiband transport. Observing a similar thickness dependence in both the transversal and the longitudinal MR in accordance with HMR theory is therefore a clear indication for the HMR. Another hint can be found in the temperature dependence. Because of the $\rho^{-n}$ scaling [Fig.~\ref{Fig_3_Kohler}(a)], the OMR in metals increases with decreasing temperature (see Appendix~D), whereas from spin (or orbital) Hall physics, a decrease in MR amplitude is typically observed \cite{velez_hanle_2016, wu_hanle_2016, sala_orbital_2023}.
	
	%<For the future observation of an orbital Hall effect in Pt, or a simultaneous observation of spin and orbital Hall effect in a material, amorphous thin films over a wide range of thicknesses with high and ideally identical resistivities should be investigated.
			
	\section{Conclusion}
	
	We investigated the MR of sputter deposited Pt thin films over a wide thickness range of \SI{4}{nm} up to \SI{120}{nm}. Careful analysis shows that the MR in all samples and directions originates from the OMR, contrary to the na\"ive expectation from the literature of Pt thin films. We find an anisotropic OMR depending on the directions of the applied field, which should be carefully considered when performing similar measurements. We propose that the HMR and the OMR coexist in samples with a spin (or orbital) Hall conductivity. Depending on the resistivity, the crystallinity and the diffusion coefficient, one or the other effect governs the measured MR. While the OMR occurs in clean, crystalline materials, the HMR dominates the MR in amorphous, sufficiently dirty samples, as there the diffusion coefficient is smaller compared to the crystalline samples of the same material. For Pt thin films we find a cutoff for a dominant HMR at approximately 5 times the bulk resistivity. In Ti we find a large HMR in the less crystalline sample which we tentatively assign to orbital origin. This suggests that to observe a spin or orbital Hanle MR in a material, it needs to feature a high resistivity, i.e., a transport that is dominated by scattering, and no crystalline order, i.e., a slow diffusion. Additionally, we observe that different reports in literature cannot be reconciled quantitatively using a simple Drude model, suggesting an insufficient understanding of the mechanism for the diffusion and dephasing of the spin and orbital accumulation.

	\section*{Acknowledgments}
	
	This work was funded by the Deutsche Forschungsgemeinschaft (DFG, German Research Foundation) via the SFB 1432, Project-ID No. 425217212 and via Project-ID No. 446571927. We also gratefully acknowledge technical support and advice by the nano.lab facility of the University of Konstanz.
	
	\section*{Appendix A: Methods}
	\label{Appendix_methods}
	
	All films from Konstanz were deposited using radio frequency (rf) magnetron sputtering in an AJA International Orion sputtering system with a base pressure better than \SI{1.5e-7}{mbar}, at room temperature and an Ar pressure of \SI{2.6e-3}{mbar} during the process. Before the sputtering, all substrates were cleaned in acetone and isopropylalcohol. Pt films were deposited onto magnesium oxide $\langle$001$\rangle$ (MgO, \textit{CrysTec}) or thermally oxidized Silicon wafers (Si/\ch{SiO_x} (\SI{1024}{nm}), \textit{Microchemicals}) at a rate of \SI{1.8}{\nano\meter\per\minute} with \SI{50}{W}, while the Ti (K$_{\mathrm{Ti}}$(30)) was sputtered with \SI{100}{W} at \SI{1.4}{\nano\meter\per\minute}. After the deposition, Hallbars [Fig.~\ref{Fig_1_Setup}(a)] with $l=$ \SI{480}{\micro m} and $w=$ \SI{50}{\micro m} were defined into the Pt films via optical lithography and subsequent Ar ion etching (Oxford Plasma Pro RIE). After contacting, the magnetotransport experiments were performed in a 3D vector magnet cryostat from Oxford Instruments, which allows an out of plane external magnetic field strength of \SI{6}{T}. 
	
\begin{figure}[t]
		\begin{center}
			\includegraphics[width=\linewidth]{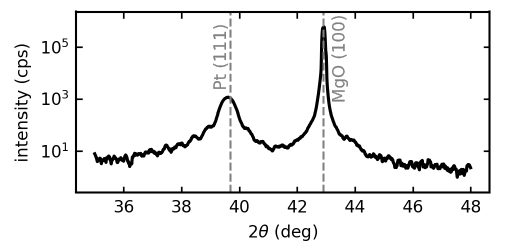}
			\caption{Symmetric $\theta$-2$\theta$ scans using Cu-K$_\alpha$ radiation on a  $\langle$001$\rangle$ oriented MgO substrate with a \SI{20}{nm} thick Pt film deposited on top. A clear peak corresponding to Pt  $\langle$111$\rangle$ can be seen at roughly \SI{39.5}{deg}, which further exhibits Laue oscillations on both sides, signalizing high crystalline quality.}
			\label{Fig_A1_XRD}
		\end{center}
	\end{figure}

	We include X-ray diffraction data of a \SI{20}{nm} thin Pt film on MgO  $\langle$001$\rangle$ [Fig.~\ref{Fig_A1_XRD}]. The measurements were performed in a Rigaku SmartLab using Cu-K$_\alpha$ radiation in $\theta$-2$\theta$ geometry. The thin film peak at \SI{39.6}{deg} corresponds to the Pt  $\langle$111$\rangle$ peak, expected for the room temperature deposition of Pt on MgO \cite{narayan_formation_1994}, which shows that the Pt film is crystalline upon deposition and highly textured along the out of plane direction.
	
	The samples from Z\"urich Z$_{\mathrm{Ti}}$(26) were prepared by direct current sputtering in a sputtering system with a base pressure of \SI{8.8e-8}{mbar}. Before the deposition the Si/\ch{SiO2} substrates were cleaned by Ar sputtering at \SI{50}{W}, \SI{0.02}{mbar} Ar pressure for \SI{60}{s}. The Ti was deposited with a power of \SI{11}{W} (\SI{40}{mA}, \SI{274}{V}) and an Ar pressure of \SI{4e-3}{mbar} at a rate of \SI{1.23}{nm\per min}. After the sputtering, \SI{8}{nm} of \ch{SiN} were deposited via rf sputtering as a capping layer. For the processing of the Hall bars and the Pt deposition parameters, refer to Sala et al. \cite{sala_orbital_2023}.
	
	\section*{Appendix B: Fuchs-Sondheimer-Model}
	\label{Appendix_FS}

	For the evaluation of the thickness dependence in Fig.~\ref{Fig_4_OMR}(a), an extended Fuchs-Sondheimer model was utilized \cite{fuchs_conductivity_1938, sondheimer_mean_1952, althammer_quantitative_2013}. The model takes a surface roughness amplitude $h$ into account to describe the increase of the resistivity towards thinner Pt films and can be described for $t_\mathrm{Pt}$ > $h$ via Eq.~\ref{Eq_FS}:
	
	\begin{equation}
		\rho_{0,\mathrm{Pt}} = \rho_{\mathrm{inf}} \left(1+\frac{3}{8(t_\mathrm{Pt}-h)} [l_\mathrm{inf}] (1-p) \right)
		\label{Eq_FS}
	\end{equation}
	
 	Here $p$ is the scattering parameter at the interface, $\rho_{\mathrm{inf}}$ the resistivity and $l_{\mathrm{inf}}$ the mean free path for an infinitely thick film, respectively. For our fits we use $p=0$ (diffusive limit), which yields the fits shown in Fig.~\ref{Fig_4_OMR}(a).
 	
 	\section*{Appendix C: Transversal resistivity}
 	
	\begin{figure}[t]
 		\begin{center}
 			\includegraphics[width=\linewidth]{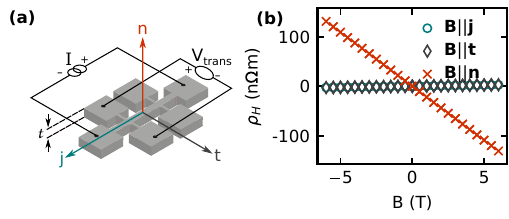}
 			\caption{(a) Experimental setup and coordinate system utilized for the transversal measurements. In the same Hallbar transversal contacts are utilized to measure the transversal voltage drop V$_\mathrm{trans}$ when sourcing a current along \textbf{j}. Like for the longitudinal geometry, the Hallbar geometry is taken into account to calculate the resistivity. (b) Transversal signal for the same \SI{8}{nm} sample seen in Fig.~\ref{Fig_1_Setup}(c). A linear dependency on the external magnetic field \textbf{B} can be seen for \textbf{B}||\textbf{n}, which can be interpreted as the ordinary Hall effect of Pt. The other two directions show no dependency as expected for the utilized geometry. For better visualization, a constant offset from imperfections of the device is removed.}
 			\label{Fig_A2_transervsesetup}
 		\end{center}
 	\end{figure}
 	
 	Next to the longitudinal measurements, transversal measurements were conducted simultaneously [Fig.~\ref{Fig_1_Setup}]. The utilized setup and geometry is seen in Fig.~\ref{Fig_A2_transervsesetup}(a). The typical transversal signal is depicted in Fig.~\ref{Fig_A2_transervsesetup}(b) for the same sample of \SI{8}{nm} Pt on MgO as seen in Fig.~\ref{Fig_1_Setup}(c).
 	
 	In this geometry, the ordinary Hall effect can typically be observed. When applying an external field \textbf{B}, a Lorentz force acts upon the electrons when the field and the direction of the current enclose a finite angle. This causes a build up of electrons perpendicular to the applied current, which can be electrically detected in our geometry for \textbf{B}||\textbf{n} [Fig.~\ref{Fig_A2_transervsesetup}]. A possible offset stemming from a spurious longitudinal contribution is removed for better visualization.
 	
 	To evaluate the transversal components, a linear fit is subtracted from the measured data along \textbf{B}||\textbf{n} in Fig.~\ref{Fig_A2_transervsesetup}. The non-linear residue is then further evaluated using Eq.~\ref{Eq_transverseHMR} as seen in the main manuscript [Fig.~\ref{Fig_6_pt_ti}]. However, when analyzing the transversal HMR contribution it becomes apparent that, depending on $\lambda$ and $D$, the curve shape appears linear within the detection limit. For a fast diffusion, this linear contribution would therefore be subtracted with the ordinary Hall effect.
	
\begin{figure}[t]
 		\begin{center}
 			\includegraphics[width=\linewidth]{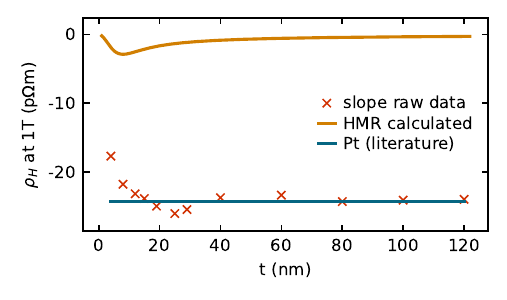}
 			\caption{Extracted transversal resistivity of the Pt samples from Fig.~\ref{Fig_2_HMR_all}. In addition, the expected signal stemming from an HMR described by Eq.~\ref{Eq_transverseHMR} with values of $\rho_0$ = \SI{249}{n\Omega m}, $\lambda_\mathrm{s}$ = \SI{2.4}{nm}, $\theta_\mathrm{s}$ = \SI{4}{\percent} and $D$~= \SI{3.1e-5}{m^2/s} is shown as the amber colored fit. The value of the ordinary Hall effect of Pt is shown in blue \cite{gehlhoff_hall-effekt_1950}. Like in YIG/Pt \cite{meyer_anomalous_2015}, a complex behavior of the Hall effect is visible for thin samples. For thicker samples, the measured transversal resistivity can coherently be explained via the ordinary Hall effect and no further contributions are observable.}
 			\label{Fig_A3_data_Hall}
 		\end{center}
 	\end{figure}

	To rule out an additional contribution next to the ordinary Hall effect, the extracted slopes of the linear fits to the data in Fig.~\ref{Fig_A2_transervsesetup} along \textbf{n} are plotted over the Pt thickness as seen in Fig.~\ref{Fig_A3_data_Hall}. Here, deviations from the expected value of \SI{-24.4}{p\Omega m} for Pt \cite{gehlhoff_hall-effekt_1950} can be seen for film thicknesses below \SI{40}{nm}. This behavior has previously been reported in yttrium iron garnet (YIG)/Pt bilayers, where thin Pt films show a complex behavior with values significantly below the expected \SI{-24.4}{p\Omega m} \cite{meyer_anomalous_2015}. Above \SI{40}{nm} of film thickness, all extracted slopes lie on the expected value of bulk Pt.
 	
 	To compare the influence of a possible spin Hanle MR, we calculate the expected transversal spin Hanle signal via Eq.~\ref{Eq_transverseHMR}. To achieve this, we again utilize $\rho_0$ = \SI{249}{n\Omega m}, $\lambda_\mathrm{s}$ = \SI{2.4}{nm}, $\theta_\mathrm{s}$ = \SI{4}{\percent} and $D$ = \SI{3.1e-5}{m^2/s}, as detailed in the main manuscript and shown in Fig.~\ref{Fig_2_HMR_all} for the longitudinal data. The resulting values are shown in the amber colored fit in Fig.~\ref{Fig_A3_data_Hall}. One observes that an additional contribution to the ordinary Hall effect would be expected for samples below \SI{20}{nm}, where the extracted data already shows a complex behavior, but no measurable effect above that thickness.
 	
	\begin{figure}[b]
 		\begin{center}
 			\includegraphics[width=\linewidth]{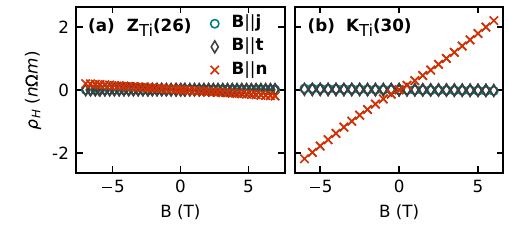}
 			\caption{Transversal resistivity as measured for the different Ti samples, Z$_\mathrm{Ti}$(26) (a) and K$_\mathrm{Ti}$(30) (b) in the same geometry as seen in Fig.~\ref{Fig_A2_transervsesetup}(a). Despite the similar thickness, a different Hall response can be observed. The crystalline sample shows a smaller amplitude with a negative sign, whereas the amorphous sample has a larger effect with a positive slope. Differences in the Hall sign have been reported previously and accounted for by temperature and crystalline differences between samples \cite{roesch_influence_1963}. Furthermore, the transversal resistivity of the amorphous sample already shows a deviation from the linear slope in the raw data, which is further supported after subtracting the ordinary Hall contribution. The subtracted data is then shown in Fig.~\ref{Fig_6_pt_ti}(j).}
 			\label{Fig_A4_Ti_data_Hall}
 		\end{center}
 	\end{figure}
 	
 	While this does not allow to exclude a spin Hanle contribution for thin samples, the transversal MR for samples above \SI{40}{nm} is conclusively described by the ordinary Hall effect. If our Pt were to show any additional HMR type effect with a diffusion length bigger than $\lambda_\mathrm{s}$ = \SI{2.4}{nm} it would appear as an addition on top of the ordinary Hall effect. We therefore conclude that for these films, no HMR type effect is present and does therefore also not exist in the longitudinal measurements and data from Fig.~\ref{Fig_2_HMR_all}. This corroborates the claim that the longitudinal MR in our Pt is caused by the ordinary MR.
 
 	To further extend our results, we included Ti samples Z$_\mathrm{Ti}$(26) and K$_\mathrm{Ti}$(30) in Fig.~\ref{Fig_6_pt_ti}(i) and (j). To analyze a possible HMR contribution, the transversal resistivity is measured along \textbf{B}||\textbf{n} as seen in Fig.~\ref{Fig_A2_transervsesetup}(a). A difference in Hall sign and amplitude can be observed for the Ti samples, where the crystalline sample Z$_\mathrm{Ti}$(26) exhibits a smaller amplitude with a negative sign [Fig.~\ref{Fig_A4_Ti_data_Hall}(a)], whereas the amorphous sample K$_\mathrm{Ti}$(30) shows a larger effect with a positive slope [Fig.~\ref{Fig_A4_Ti_data_Hall}(b)]. Differences in the Hall slope and sign have previously been reported for Ti where they were traced back to a temperature and crystal orientation dependent Hall effect \cite{roesch_influence_1963}. As both measurements where performed at \SI{300}{K}, the temperature can be excluded. However, as Fig.~\ref{Fig_6_pt_ti}(f) depicts, large differences in the crystal properties can be observed. We therefore suggest the different crystalline structures of the samples as the main reason for the contrasting slopes and amplitudes.
 	
 	Like for Pt [Fig.~\ref{Fig_6_pt_ti}(d),(e)], a non-linear residual can be found in the amorphous K$_\mathrm{Ti}$(30) sample [Fig.~\ref{Fig_A4_Ti_data_Hall}(b)], whereas no additional contribution is found in Z$_\mathrm{Ti}$(26) [Fig.~\ref{Fig_A4_Ti_data_Hall}(a)]. This becomes evident after subtracting the linear component of the ordinary Hall effect from the data, as described in the main manuscript [Fig.~\ref{Fig_6_pt_ti}(i) and(j)].
 	
 	\begin{figure}[t]
	 	\begin{center}
	 		\includegraphics[width=\linewidth]{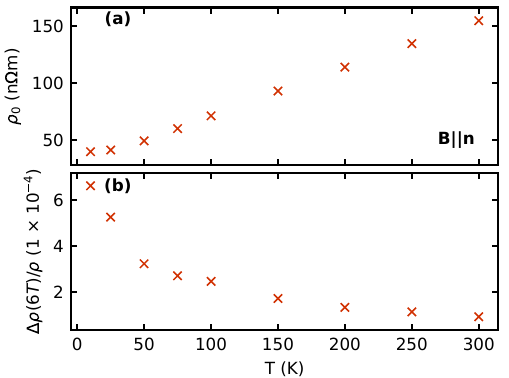}
	 		\caption{(a) Temperature dependence of the resistivity and (b) the longitudinal MR for \textbf{B}||\textbf{n} at \SI{6}{T}. The resistivity decreases with temperature as expected for Pt \cite{velez_hanle_2016, wu_hanle_2016}. The MR on the other hand increases with temperature, signaling again a scaling with $\rho^{-n}$ and the OMR being the main contributor. For spin Hall type effects, a decrease of the MR is usually reported and expected \cite{velez_hanle_2016, wu_hanle_2016, sala_orbital_2023}.}
	 		\label{Fig_A5_Tdep}
	 	\end{center}
	 \end{figure}
 	\section*{Appendix D: Temperature dependence of the longitudinal MR} 
	
	To further support the claim that our MR is dominated by the ordinary MR, we include the temperature dependence of an exemplary set for \textbf{B}||\textbf{n}. Figure~\ref{Fig_A5_Tdep}(a) shows the resistivity and Fig.~\ref{Fig_A5_Tdep}(b) the longitudinal MR versus temperature. As expected for the ordinary MR, the MR increases with lower temperature as the resistivity decreases. Fewer scattering processes increase the effect of the magnetic field on the charge carriers and in turn increase the amplitude of the OMR \cite{pippard_magnetoresistance_1989}. This further corroborates the $\rho^{-n}$ scaling and further rules out the HMR, as for spin Hall type effects a decrease in MR with decreasing temperature is expected, as with fewer scattering processes, fewer spin (or orbital) dependent scattering is possible \cite{velez_hanle_2016, wu_hanle_2016, sala_orbital_2023}.
	%\section{References}
	%\bibliographystyle{apsrev4-2}
	\bibliography{MyLibrary.bib}% Produces the bibliography via BibTeX.
	
\end{document}